\documentclass[10pt]{article}
\usepackage{titlesec}
\usepackage{amsmath,esdiff}
\usepackage{amssymb}     
\usepackage{authblk}
\usepackage[normalem]{ulem}
\RequirePackage{bbold,bm}
\usepackage{bbold,bm}
\usepackage{geometry}
\usepackage{BOONDOX-cal}
\usepackage{setspace}
\usepackage{cite}
\usepackage[colorlinks=true,linkcolor=blue,citecolor=red,urlcolor=red]{hyperref}%
\usepackage{mathrsfs}
\usepackage{hyperref}
\usepackage{graphicx}
\usepackage{flushend}
\usepackage{cancel}
\usepackage{array,esint}
\usepackage[most]{tcolorbox}
\titleformat{\section}{\fontsize{12}{12}\bfseries}{\thesection}{1em}{}
\graphicspath{}
\twocolumn
\begin{document}
\twocolumn[\begin{@twocolumnfalse}
\title{\textbf{Near horizon aspects (and beyond) of acceleration radiation of an atom falling into a large class of static spherically symmetric black hole geometries}}
\author{\textbf{$\mathbf{Soham}$ $\mathbf{Sen}^\dagger$, $\mathbf{Rituparna}$ $\mathbf{Mandal}^*$ and $\mathbf{Sunandan}$ $\mathbf{Gangopadhyay}^\ddagger$}}
\affil{{Department of Theoretical Sciences}\\
{S.N. Bose National Centre for Basic Sciences}\\
{JD Block, Sector III, Salt Lake, Kolkata 700 106, India}}
\date{}
\maketitle
\begin{abstract}
\noindent The near horizon aspects (and beyond) of a black hole metric, which belongs to a large class of static spherically symmetric black holes, is considered here. It has been realized recently that an atom falling into a black hole leads to the generation of acceleration radiation through virtual transitions. In recent studies it has been argued that this acceleration radiation can be understood from the near horizon physics of the black hole. The near horizon approximation leads to conformal symmetry in the problem. We go beyond the near horizon approximation in our analysis. This breaks the conformal symmetry associated with the near horizon physics of the black hole geometry. We observe that even without the consideration of the conformal symmetry, the modified equivalence relation holds. Further, our analysis reveals that the probability of virtual transition retains its Planck like form with the amplitude getting modified due to the beyond near horizon approximation. For the next part of our analysis, we have observed the horizon brightened acceleration radiation entropy (HBAR) for a Garfinkle-Horowitz-Strominger (GHS) black hole. We observe that the HBAR entropy misses out on quantum gravity like corrections while considering the conformal case. However, such corrections emerge when the conformal symmetry gets broken in the beyond near horizon analysis.
\end{abstract}
\end{@twocolumnfalse}]
\section*{Introduction}
\noindent \footnote{{\quad}\\
{${}^\dagger$soham.sen@bose.res.in, sensohomhary@gmail.com}\\
{${}^*$drimit.ritu@gmail.com}\\
{${}^\ddagger$sunandan.gangopadhyay@gmail.com}}
With the advent of general theory of relativity\cite{Einstein15, Einstein16}, the concept of the black hole arose as an exact solution to the non-linear field equations of general relativity. Stephen Hawking, in his papers\cite{Hawking,Hawking2,Hawking3}, showed that a black hole emits radiation while considering quantum effects in curved space-time. The efforts to unite gravity, geometry, and thermodynamics led to seminal works like black hole thermodynamics\cite{Hawking,Hawking2,Hawking3,Bekenstein,Bekenstein2}, Hawking radiation\cite{Hawking,Hawking2,Hawking3}, particle emission from black holes\cite{Page,Page2,Page3}, the Unruh effect\cite{Unruh}, and acceleration radiation\cite{Fulling21,Davies,DeWitt,Unruh2, Muller,Vanzella,Higuichi, Fulling2,Ordonez1,Ordonez2,Ordonez3,Ordonez4,OTM}. The Hawking radiation from a black hole can be directly related to the Unruh effect which is nothing but the detection of particles by an accelerated detector in an inertial vacuum. In recent times, there has been an upsurge in the efforts to analytically explain the combined effects of quantum optics and black hole physics\cite{Fulling2,OTM,Weiss,Philbin}. In this scenario, the two-level atom behaves as a detector. In several recent works, it has been shown that an atom falling into a black hole with an arbitrary metric feels the effect of thermal radiation similar to that of Hawking radiation\cite{Fulling2,OTM}. The entropy of radiation emitted in this process is termed as horizon brightened acceleration radiation entropy (HBAR) which is a tool to understand the relation between atom optics and general relativity as mentioned above\cite{Fulling2}. A few works have also focused on the effects of conformal symmetry in analyzing the near horizon aspects of a black hole\cite{Conformal1,Conformal2,Ordonez1,Ordonez2,Ordonez3,Ordonez4}. In these works, the main focus is on the near horizon aspects of a generalized $D$-dimensional Schwarzschild black hole and its relation with Hawking radiation in case of the existence of conformal symmetry. In this work, we consider a large class of $D$-dimensional static spherically symmetric black hole spacetimes and study the near horizon aspects (and beyond) of acceleration radiation of an atom falling into this geometrical setup. In particular, we go beyond the near horizon expansion of the black hole metric and study its consequences in the acceleration radiation of the atom. The results that we obtain are important in its own right because they capture the beyond near horizon aspects of the black hole in the problem of virtual transition and HBAR entropy.

\noindent A generalized spacetime geometry arises in case of low energy string theory\cite{ACV,KOPAPR} which are quantum gravity theories that combine all the fundamental forces of nature into one single unified theory\footnote{Other attempts to combine gravity with quantum mechanics are loop quantum gravity\cite{ROVELLI,CARLIP} and noncommutative geometry\cite{Girelli}.}. 
The Garfinkle-Horowitz-Strominger (GHS) black hole is a solution of a low energy string theoretic action\cite{GHS1,GHS2,SG0}. It reads
\begin{equation}\label{1.2}
ds^2=-f(r)dt^2+g(r)^{-1}dr^2+r^2d\Omega^2~.
\end{equation}
The analytical forms of $f(r)$ and $g(r)$ reads
\begin{align}
f(r)&=\left(1-\frac{2Me^{\phi_0}}{r}\right)\left(1-\frac{Q^2e^{3\phi_0}}{Mr}\right)^{-1}~,\label{1.3}\\
g(r)&=\left(1-\frac{2Me^{\phi_0}}{r}\right)\left(1-\frac{Q^2e^{3\phi_0}}{Mr}\right)\label{1.4}
\end{align}
where $\phi_0$ is the asymptotic constant value of the dilaton field.

\noindent Inspired by the earlier works in this field and the previously discussed generic black model (GHS black hole), we have considered here a $D$-dimensional static spherically symmetric black hole metric. Our aim is to calculate the excitation probability of a virtual transition due to an atom falling into the event horizon of such a $D$-dimensional static spherically symmetric black hole.  Our next goal is to check for the equivalence principle for such a scenario with and without the consideration of conformal symmetry. The conformal symmetry in the near horizon avatar of the Klein Gordon equation can be broken by keeping higher-order terms in the near horizon expansion of the metric. We shall look precisely into the role played by the higher-order terms. There were several attempts to give an alternative explanation of the equivalence principle\cite{Rohrlich,Vallisneri, Singleton,Singleton2,Fulling,Fulling0} along with the recent consideration of generalized uncertainty relation for an atom-mirror system\cite{SG}.  It was observed in \cite{Fulling2} that an atom falling into a black hole emits a similar spectrum to that of a mirror which is accelerating uniformly with respect to a fixed atom. We will investigate this modified equivalence relation in the most general setup possible. For the next part of our work, we will calculate the transition and absorption rates for the GHS black hole geometry from the analytical expressions derived for the generic metric with an event horizon at $r=r_+$ such that $f(r_+)=g(r_+)=0$. Using the density matrix formalism from atom-optics, we shall calculate the horizon brightened acceleration radiation (HBAR) entropy for the GHS black hole.

\noindent The paper is organized as follows. In section 1, we provided a basic description of the near horizon approximation and obtained the Rindler form of the static spherically symmetric black hole geometry. In section 2, we have found a solution to the covariant Klein-Gordon field equation with and without considering conformal symmetry. In section 3, we have obtained the transition probabilities due to virtual transition for an atom falling  
into the event horizon of a static spherically symmetric black hole metric. In section 4, we obtained the HBAR entropy for the GHS black hole. We conclude in section 5.


\section{The generalized and static black hole geometry and the basic formalism}
The metric describing a large class of static spherically symmetric black holes is given as follows (in natural units)
\begin{equation}\label{1.5}
ds^2=-f(r)dt^2+g(r)^{-1}dr^2+r^2d\Omega^2_{D-2}
\end{equation}
where, $D$ is the number of spacetime dimensions with the condition $D\geq 4$. Here, $r_+$ is the horizon radius such that
\begin{equation}\label{1.6}
f(r_+)=g(r_+)=0~.
\end{equation}
Under a near horizon expansion (keeping terms upto first order in $(r-r_+)$), $f(r)$ and $g(r)$ (satisfying eq.(\ref{1.6})) takes the following form
\begin{equation}\label{1.7}
f(r)\cong(r-r_+)f'(r_+)~,~g(r)\cong(r-r_+)g'(r_+)
\end{equation} 
where $'$ denotes derivative with respect to $r$. Note that while writing the covariant Klein-Gordon equation, we will use higher order corrections in the near horizon approximation. We now define a transformation of coordinates as follows
\begin{equation}\label{1.8}
\rho=2\sqrt{\frac{r-r_+}{g'(r_+)}}~.
\end{equation}
Using eq.(s)(\ref{1.7},\ref{1.8}) in eq.(\ref{1.5}), we obtain the following form of the metric
\begin{equation}\label{1.9}
ds^2=-\frac{\rho^2f'(r_+)g'(r_+)}{4}dt^2+d\rho^2+r^2d\Omega^2_{D-2}~.
\end{equation}
This particular form of the metric given in eq.(1.9) is also known as the Rindler form of the metric in $D$-spacetime dimensions. Following earlier literature\cite{OTM}, we can compute the uniform acceleration corresponding to curves of constant $\rho$ as follows
\begin{equation}\label{1.10}
a=\frac{1}{\rho}=\frac{1}{2}\sqrt{\frac{g'(r_+)}{r-r_+}}~.
\end{equation}
We shall use this acceleration to check for the validity of the modified equivalence relation in this generic set up.
\section{Solution of the covariant Klein Gordon field equation}
The covariant Klein-Gordon quantum field equation for a  scalar field with rest mass $m_0$ reads
\begin{equation}\label{1.11}
\frac{1}{\sqrt{-g}}\partial_\mu(\sqrt{-g}g^{\mu\nu}\partial_\nu\Psi)-m_0^2\Psi=0
\end{equation}
where the mode expansion of the scalar field $\Psi$ reads
\begin{equation}\label{1.12}
\Psi(t,r,\Omega)=\sum\limits_{n,l,m}\left[b_{nlm}\psi_{nlm}(t,r,\Omega)+h.c.\right]
\end{equation}
with $b_{nlm}$ being the field annihilation operator. We can now use a separation of variables for $\psi_{nlm}$ as follows
\begin{equation}\label{1.13}
\psi_{nlm}=\zeta(r)u_{nl}(r)Y_{lm}(\Omega)e^{-i\nu_{nl}t}
\end{equation}
with $\nu_{nl}$ being the frequency of the scalar mode and $Y_{lm}(\Omega)$ being the spherical harmonics. Note that $\psi_{nlm}$ satisfies the classical Klein-Gordon field equation. We take a form of $\zeta(r)$ as follows
\begin{equation}\label{1.14}
\begin{split}
\zeta(r)&=\exp\left[-\frac{1}{2}\int dr\left(\frac{f'(r)}{2f(r)}+\frac{g'(r)}{2g(r)}+\frac{D-2}{r}\right)\right]\\
&=\left(f(r)g(r)\right)^{-\frac{1}{4}}r^{-\left(\frac{D}{2}-1\right)}~.
\end{split}
\end{equation}
Eq.(\ref{1.11}) can be recast in the following form
\begin{equation}\label{1.15}
\begin{split}
&-\frac{\partial_t^2\Psi}{f(r)}+\left(\frac{D-2}{r}+\frac{\partial_rf(r)}{2f(r)}+\frac{\partial_rg(r)}{2g(r)}\right)g(r)\partial_r\Psi\\&
+g(r)\partial_r^2\Psi+\frac{1}{r^2}\Delta^{D-2}_{\Gamma}\Psi-m^2\Psi=0
\end{split}
\end{equation}
where $\Gamma$ is the metric on the unit $(D-2)$-dimensional sphere, $S^{D-2}$. Using eq.(s)(\ref{1.13},\ref{1.14}) in eq.(\ref{1.15}), we obtain the following radial equation
\begin{equation}\label{1.16}
\begin{split}
&u_{nl}''(r)+\biggr[-\left(\frac{f''(r)}{4f(r)}+\frac{g''(r)}{4g(r)}+\frac{D-2}{4r}\left(\frac{g'(r)}{g(r)}+\frac{f'(r)}{f(r)}\right)\right)\\
&+\left(\frac{\nu_{nl}^2}{f(r)g(r)}+\frac{3}{8}\left(\frac{g'^2(r)}{2g^2(r)}+\frac{f'^2(r)}{2f^2(r)}-\frac{f'(r)g'(r)}{3f(r)g(r)}\right)\right)\\
&-\frac{\alpha+r^2\beta}{r^2g(r)}+\biggr(\left(\frac{1}{g(r)}-1\right)\frac{(D-3)^2}{4r^2}+\frac{1}{4r^2}\biggr)\biggr]u_{nl}(r)=0
\end{split}
\end{equation}
where $\beta=m_0^2$ and $\alpha=\left(l+\frac{D-3}{2}\right)^2$. We shall now carry out our analysis by going beyond the near horizon (beyond NH) approximation. The forms of $f(r)$ and its higher order derivatives in this beyond NH formalism are given by
\begin{align}
f(r)&\cong (r-r_+)f'(r_+)+\frac{(r-r_+)^2}{2}f''(r_+)\nonumber\\
&+\mathcal{O}((r-r_+)^3),\label{1.16a}\\
f'(r)&\sim f'(r_+)+(r-r_+)f''(r_+)~,\label{1.16b}\\
f''(r)&\sim f''(r_+)\label{1.16c}~.
\end{align}
$g(r)$, $g'(r)$ and $g''(r)$ has similar structure to that of $f(r)$, $f'(r)$ and $f''(r)$. For notational simplicity we will be using $r-r_+=y$, $u_{nl}(r)=u$, $\nu_{nl}=\nu$, $f'(r_+)=f'_+$ and $f''(r_+)=f"_+$ (similarly for $g'(r_+)$ and $g''(r_+)$). In this beyond near horizon approximation eq.(\ref{1.16}) takes the following form
\begin{equation}\label{1.17}
\begin{split}
&u''+\left(\frac{\nu^2}{f'_+g'_+}+\frac{1}{4}\right)\frac{u}{y^2}-\biggr[\left(\frac{\nu^2}{f'_+g'_+}+\frac{1}{4}\right)\left(\frac{f''_+}{2f'_+}+\frac{g''_+}{2g'_+}\right)\\&+\frac{(D-2)}{2r_+}+\frac{\alpha+r_+^2\beta}{r_+^2g'_+}-\frac{(D-3)^2}{4g'_+r_+^2}\biggr]\frac{u}{y}=0~.
\end{split}
\end{equation}
We can rewrite eq.(\ref{1.17}) in a more simpler form as follows
\begin{equation}\label{1.18}
y^2u''-\kappa y u+\left(\frac{\nu^2}{f'_+g'_+}+\frac{1}{4}\right)u=0.
\end{equation}
The form of $\kappa$ in the above equation is given as
\begin{equation}\label{1.19}
\begin{split}
\kappa&=\left(\frac{\nu^2}{f'_+g'_+}+\frac{1}{4}\right)\left(\frac{f''_+}{2f'_+}+\frac{g''_+}{2g'_+}\right)+\frac{(D-2)}{2r_+}\\&+\frac{\alpha+r_+^2\beta}{r_+^2g'_+}-\frac{(D-3)^2}{4g'_+r_+^2}~.
\end{split}
\end{equation}
With the general form of the equation in hand, we can now proceed to obtain the solution of the radial equation. As $r-r_+=y$ is a very small quantity, $\mathcal{O}\left(\frac{1}{y^2}\right)$ term is the most dominant term in eq.(\ref{1.17}). If we neglect the $\mathcal{O}\left(\frac{1}{y}\right)$ term in eq.(\ref{1.17}), we obtain the following equation
\begin{equation}\label{1.20}
y^2u''(y)+\left(\frac{\nu^2}{f'_+g'_+}+\frac{1}{4}\right)u(y)=0~.
\end{equation}
This equation has an interesting property. It exhibits an asymptotic conformal symmetry. This can be seen by taking an ansatz of the form 
\begin{equation}\label{1.20a}
u(y)=y^q~.
\end{equation}
Putting this ansatz back in eq.(\ref{1.20}), we obtain a quadratic equation in $q$ given as
\begin{equation}\label{1.20b}
q^2-q+\frac{\nu^2}{f'_+g'_+}+\frac{1}{4}=0~.
\end{equation}
 Solving the above equation, we obtain the analytical form of $q$ as
\begin{equation}
q=\frac{1}{2}\pm i\frac{\nu}{\sqrt{f'_+g'_+}}~.
\end{equation} 
The solution of eq.(\ref{1.20}) can be recast in the following form
\begin{equation}\label{1.21}
u(y)=y^{\frac{1}{2}\pm i\frac{\nu}{\sqrt{f'_+g'_+}}}~.
\end{equation}
The outgoing part of the solution in eq.(\ref{1.21}) is given as 
\begin{equation}\label{1.22}
u(y)=y^{\frac{1}{2}+i\frac{\nu}{\sqrt{f'_+g'_+}}}~.
\end{equation} 
The fact that scaling solutions are allowed for eq.(\ref{1.20}) validates the conformal symmetry in the problem. 
Further, from eq.(\ref{1.20}), we can observe that the near horizon physics corresponds to a one-dimensional effective Hamiltonian
\begin{equation}\label{1.22a}
H=p_y^2-\frac{\lambda}{y^2}
\end{equation}
where $\lambda=\left(\frac{\nu^2}{f'_+g'_+}+\frac{1}{4}\right)>0$. This is the representation of the well known long-range conformal quantum mechanics\cite{Ordonez5}.

\noindent The position and time dependent part of the solution, therefore is given by
\begin{equation}\label{1.23}
\begin{split}
\psi(r,t)&=e^{-i\nu t}\zeta(r) u(r)=\mathcal{K}e^{-i\nu t}y^{i\frac{\nu}{\sqrt{f'_+g'_+}}}
\end{split}
\end{equation}
with $\mathcal{K}=r_+^{\frac{D}{2}-1}(f'_+g'_+)^{-\frac{1}{4}}$. We will now investigate the case where we now include the  $\mathcal{O}(1/y)$ term in eq.(\ref{1.18}). Note that the ansatz in eq.(\ref{1.20a}) will not go through now since eq.(\ref{1.18}) does not have the conformal symmetry. We assume a solution of the form given by
\begin{equation}\label{1.24}
u(y)=\sum\limits_{l=0}^\infty A_l y^{s+l}
\end{equation}
where $s$ is a constant. Now substituting eq.(\ref{1.24}) back in eq.(\ref{1.18}), we get the following relation
\begin{equation}\label{1.25}
\sum\limits_{j=0}^{\infty}A_jy^{j+s}\left[(s+j)(s+j-1)-\kappa y+\frac{\nu^2}{f'_+g'_+}+\frac{1}{4}\right]=0~.
\end{equation}
We will now compare the coefficients for equal powers of $y$.
Equating the coefficient of the $y^s$ term to zero, we get the following quadratic equation for $s$
\begin{equation}\label{1.26}
s^2-s+\frac{\nu^2}{f'_+g'_+}+\frac{1}{4}=0
\end{equation} 
which has a solution given by
\begin{equation}\label{1.27}
s=\frac{1}{2}\pm i\frac{\nu}{\sqrt{f'_+g'_+}}~.
\end{equation}
The higher order equations in $y$ yield a recursion relation among the coefficients as follows
\begin{equation}\label{1.28}
A_n=\frac{\kappa A_{n-1}}{(n+s)(n+s-1)+\frac{\nu^2}{f'_+g'_+}+\frac{1}{4}}~;~\{n=1,2,3,\ldots\}~. 
\end{equation}
For $s=\frac{1}{2}+\frac{i\nu}{\sqrt{f'_+g'_+}}$, the generalized form of the coefficient $A_n$ with arbitrary value of $n$ is given as
\begin{equation}\label{1.29}
A_n=\frac{\kappa^n A_0}{n!\prod\limits_{l=1}^{n}\left(l+i\frac{2\nu}{\sqrt{f'_+g'_+}}\right)}=\frac{\kappa^n A_0\Gamma\left(1+i\frac{2\nu}{\sqrt{f'_+g'_+}}\right)}{n!\Gamma\left(n+1+i\frac{2\nu}{\sqrt{f'_+g'_+}}\right)}~.
\end{equation}
For simplicity we will be using, $A_0=1$. Neglecting higher order terms, the solution in eq.(\ref{1.24}) takes the following form
\begin{equation}\label{1.30}
\begin{split}
u(y)&=y^{\frac{1}{2}+i\frac{\nu}{\sqrt{f'_+g'_+}}}\sum\limits_{n=0}^\infty\frac{y^n\kappa^n \Gamma\left(1+i\frac{2\nu}{\sqrt{f'_+g'_+}}\right)}{n!\Gamma\left(n+1+i\frac{2\nu}{\sqrt{f'_+g'_+}}\right)}~.
\end{split}
\end{equation}
Therefore, the position and time dependent part of the classical scalar field solution takes the following form
\begin{equation}\label{1.31}
\psi_\kappa(r,t)\sim e^{-i\nu t}y^{i\frac{\nu}{\sqrt{f'_+g'_+}}}\sum\limits_{n=0}^\infty\frac{y^n\kappa^n \Gamma\left(1+i\frac{2\nu}{\sqrt{f'_+g'_+}}\right)}{n!\Gamma\left(n+1+i\frac{2\nu}{\sqrt{f'_+g'_+}}\right)}~.
\end{equation}
In the next section, we will compute the transition probability due to virtual transition for an atom falling into the event horizon of a black hole which belongs to a large class of static spherically symmetric geometries.
\section{Calculating the transition probabilities}
To calculate the transition probability, we need the atom-trajectory at first. We will simplify our calculations by restricting to a four dimensional case. In terms of the Killing vectors $\bm{\varrho}=\partial_t$, $\bm{\varsigma}=\partial_\phi$ and spacetime velocity $\bm{u}$, we can define two conserved quantities as follows
\begin{equation}\label{1.32}
\frac{E}{m}=-\bm{\varrho}\cdot\bm{u}=f(r)\frac{dt}{d\tau}~,~\frac{L}{m}=\bm{\varsigma}\cdot \bm{u}=r^2\frac{d\phi}{d\tau}
\end{equation}
where $\mathcal{e}=\frac{E}{m}$ is the relativistic energy per unit rest mass of the atom and $\mathcal{l}=\frac{L}{m}$ is the angular momentum per unit rest mass of the atom. In terms of the above conserved quantities, the spacetime geometry (eq.(\ref{1.5}) for $D=4$) can be recast in the following form
\begin{equation}\label{1.32a}
\begin{split}
1&=f(r)\left(\frac{dt}{d\tau}\right)^2-\frac{1}{g(r)}\left(\frac{dr}{d\tau}\right)^2-\frac{\mathcal{l}^2}{r^2}\\
\left(1+\frac{\mathcal{l}^2}{r^2}\right)&=\frac{\mathcal{e}^2}{f(r)}-\frac{1}{g(r)}\left(\frac{dr}{d\tau}\right)^2\\
d\tau&=-\sqrt{\frac{f(r)}{g(r)}}\frac{dr}{\sqrt{\mathcal{e}^2-f(r)\left(1+\frac{\mathcal{l}^2}{r^2}\right)}}\\
\implies \tau(r)&=-\int\sqrt{\frac{f(r)}{g(r)}}\frac{dr}{\sqrt{\mathcal{e}^2-f(r)\left(1+\frac{\mathcal{l}^2}{r^2}\right)}} ~.
\end{split}
\end{equation} 
Using this relation in eq.(\ref{1.32}), we can obtain the form of $t(r)$ as follows
\begin{equation}\label{1.32b}
\begin{split}
\mathcal{e}&=f(r)\frac{dt}{d\tau}\\
dt&=-\frac{\mathcal{e}}{\sqrt{f(r)g(r)}}\frac{dr}{\sqrt{\mathcal{e}^2-f(r)\left(1+\frac{\mathcal{l}^2}{r^2}\right)}}\\
\implies t(r)&=-\int\frac{\mathcal{e}}{\sqrt{f(r)g(r)}}\frac{dr}{\sqrt{\mathcal{e}^2-f(r)\left(1+\frac{\mathcal{l}^2}{r^2}\right)}}~.
\end{split}
\end{equation}
The proper time $\tau$ and $t$ which give the atom trajectories can therefore be written as follows
\begin{align}
\tau(r)&=-\int\frac{1}{\mathcal{e}}\sqrt{\frac{f(r)}{g(r)}}\frac{dr}{\sqrt{1-\frac{f(r)}{\mathcal{e}^2}\left(1+\frac{\mathcal{l}^2}{r^2}\right)}}\label{1.33}~,\\
t(r)&=-\int\frac{1}{\sqrt{f(r)g(r)}} \frac{dr}{\sqrt{1-\frac{f(r)}{\mathcal{e}^2}\left(1+\frac{\mathcal{l}^2}{r^2}\right)}}\label{1.34}~.
\end{align}
Applying the beyond near horizon approximation along with the consideration $\mathcal{l}=0$, we will now calculate the atom trajectories upto $\mathcal{O}(y)$.  The forms of $\tau$ and $t$ upto $\mathcal{O}(y)$ are given by
\begin{align}
\tau&=-\frac{1}{\mathcal{e}}\sqrt{\frac{f'_+}{g'_+}}y+\mathcal{C}_1\label{1.35}~,\\
t&=-\frac{1}{\sqrt{f'_+g'_+}}\left(\ln y+y\left(\frac{f'_+}{2\mathcal{e}^2}-\frac{1}{4}\left(\frac{f''_+}{f'_+}+\frac{g''_+}{g'_+}\right)\right)\right)+\mathcal{C}_2\label{1.36}
\end{align}
where $\mathcal{C}_1~\text{and}~\mathcal{C}_2$ are integration constants. We will at first calculate the transition probability for the case when the $\mathcal{O}(1/y)$ term is kept in the analysis. The atom field interaction Hamiltonian has the form given as follows 
\begin{equation}\label{1.37}
\hat{\mathcal{H}}_I=\hbar\mathcal{G}\left[\hat{b}_\nu\psi_\nu(t(r),r(\tau))+h.c.\right]\left[\hat{\xi}e^{-i\varpi \tau}+h.c.\right]
\end{equation}
where $\mathcal{G}$ is the atom-field coupling constant, $\hat{b}_\nu$ is the photon annihilation operator, $\varpi$ is the atom transition frequency and $\hat{\xi}$ is given by $\left|g\right>\left<e\right|$ with $\left|e\right>$ and $\left|g\right>$ denoting the excited and ground states of the two-level atom. 
The transition probability concerning eq.(\ref{1.23}) reads
\begin{equation}\label{1.38}
\begin{split}
P_{\kappa}&=\frac{1}{\hbar^2}\left|\int d\tau \left<1_\nu,e\right|\hat{\mathcal{H}}_I\left|0_\nu,g\right>\right|^2\\
&=\mathcal{G}^2\left|\int d\tau~\psi_\kappa^*(r,t)~e^{i\varpi\tau}\right|^2~.
\end{split}
\end{equation}
We observe that the contribution to the transition probability in eq.(\ref{1.38}) is due to the counter-rotating terms in eq.(\ref{1.37}). To proceed further, we make a simple change in variables as given by $y=\frac{\mathcal{e}x}{\varpi}$, where $x,\nu\ll \varpi$\cite{Fulling2}. Substituting the form of $\psi_\kappa(r,t)$ from eq.(\ref{1.31}) in the above equation, we get
\begin{equation}\label{1.39}
\begin{split}
P_\kappa&\cong\frac{\mathcal{G}^2}{\varpi^2}\biggr|\sqrt{\frac{f'_+}{g'_+}}\int_0^{\infty}\biggr(1+\frac{\kappa \mathcal{e}x}{\varpi\biggr(1-\frac{2 i\nu}{\sqrt{f'_+g'_+}}\biggr)}+\frac{\mathcal{e}x}{\varpi}\biggr(\frac{f'_+}{2\mathcal{e}^2}\\&+\frac{1}{4}\left(\frac{f''_+}{f'_+}-\frac{g''_+}{g'_+}\right)\biggr)\biggr)x^{-\frac{2i\nu}{\sqrt{f'_+g'_+}}}\\
&\times e^{-ix\left[\sqrt{\frac{f'_+}{g'_+}}+\frac{\mathcal{e}\nu}{\varpi}\frac{1}{\sqrt{f'_+g'_+}}\left(\frac{f'_+}{2\mathcal{e}^2}-\frac{1}{4}\left(\frac{f''_+}{f'_+}+\frac{g''_+}{g'_+}\right)\right)\right]} \biggr|^2~.
\end{split}
\end{equation}
We shall now make another change of variables as follows
\begin{equation}\label{1.40}
xf'_+=\chi\sqrt{f'_+g'_+}~.
\end{equation}
Using eq.(\ref{1.40}) in eq.(\ref{1.39}), we obtain the form of the probability as follows
\begin{equation}\label{1.41}
\begin{split}
P_\kappa&=\frac{\mathcal{G}^2}{\varpi^2}\biggr|\int d\chi\biggr[1+\frac{\mathcal{e}\chi}{\varpi}\sqrt{\frac{g'_+}{f'_+}}\biggr[\mu+\frac{f'_+}{2\mathcal{e}^2}+\frac{1}{4}\left[\frac{f''_+}{f'_+}-\frac{g''_+}{g'_+}\right]\biggr]\\&
+\frac{2i\mathcal{e}\mu\nu\chi}{\varpi f'_+}\biggr]e^{-i\chi\left(1+\frac{\nu\mathcal{e}}{\varpi f'_+}\left(\frac{f'_+}{2\mathcal{e}}-\frac{1}{4}(\frac{f''_+}{f'_+}+\frac{g''_+}{g'_+})\right)\right)}\chi^{-\frac{2i\nu}{\sqrt{f'_+g'_+}}}\biggr|^2
\end{split}
\end{equation}
where $\mu=\frac{\kappa}{1+\frac{4\nu^2}{f'_+g'_+}}$. 

\noindent To obtain the final form of the probability, we will introduce a change of variables given by
\begin{equation}\label{1.41a}
\zeta=\chi\mathcal{B}
\end{equation}
where
\begin{equation}\label{1.41b}
\mathcal{B}=\left(1+\frac{\nu\mathcal{e}}{\varpi f'_+}\left(\frac{f'_+}{2\mathcal{e}}-\frac{1}{4}\left(\frac{f''_+}{f'_+}+\frac{g''_+}{g'_+}\right)\right)\right)~.
\end{equation}
Using, the coordinate transformation in eq.(\ref{1.41a}), we obtain the final form of the excitation probability as
\begin{equation}\label{1.41c}
\begin{split}
P_\kappa&=\frac{\mathcal{G}^2}{\mathcal{B}^2\varpi^2}\biggr|\int_0^\infty d\zeta~ e^{-i\zeta}~\zeta^{-\frac{2i\nu}{\sqrt{f'_+g'_+}}}\biggr[1+\frac{\mathcal{e}\zeta}{\mathcal{B}\varpi}\sqrt{\frac{g'_+}{f'_+}}\biggr[\mu+\\&\frac{f'_+}{2\mathcal{e}^2}+\frac{1}{4}\left[\frac{f''_+}{f'_+}-\frac{g''_+}{g'_+}\right]\biggr]
+\frac{2i\mathcal{e}\mu\nu\zeta}{\mathcal{B}\varpi f'_+}\biggr]\biggr|^2~.
\end{split}
\end{equation}\label{1.41f}
Next we will use another quantity given by
\begin{equation}
\gamma=\mu+\frac{f'_+}{2\mathcal{e}^2}+\frac{1}{4}\left[\frac{f''_+}{f'_+}-\frac{g''_+}{g'_+}\right].
\end{equation}
As $\nu\ll\varpi$, we will keep only up to $\mathcal{O}(\frac{\nu}{\varpi})$ terms in our calculation. Hence, we can express $\frac{1}{\mathcal{B}}$ as
\begin{equation}\label{1.41h}
\frac{1}{\mathcal{B}}\cong\left(1-\frac{\nu\mathcal{e}}{\varpi f'_+}\left(\frac{f'_+}{2\mathcal{e}}-\frac{1}{4}\left(\frac{f''_+}{f'_+}+\frac{g''_+}{g'_+}\right)\right)\right)~.
\end{equation}
Up to $\mathcal{O}(\frac{\nu}{\varpi})$, $\frac{1}{\mathcal{B}^2}$ can similarly be expressed as
\begin{equation}\label{1.41e}
\frac{1}{\mathcal{B}^2}\cong 1-\frac{2\nu\mathcal{e}}{\varpi f'_+}\left(\frac{f'_+}{2\mathcal{e}}-\frac{1}{4}\left(\frac{f''_+}{f'_+}+\frac{g''_+}{g'_+}\right)\right)~.
\end{equation} 
Using the form of $\gamma$ in eq.(\ref{1.41f}), we can express the final form of the probability in a much simpler form given by
\begin{equation}\label{1.42}
\begin{split}
P_\kappa&=\frac{\mathcal{G}^2}{\varpi^2\mathcal{B}^2}\biggr|\int_0^\infty d\zeta\biggr[1+\frac{\mathcal{e}\zeta\gamma}{\mathcal{B}\varpi}\sqrt{\frac{g'_+}{f'_+}}+\frac{2i\mathcal{e}\mu\nu\zeta}{\mathcal{B}\varpi f'_+}\biggr]\\&\times e^{-i\zeta}\zeta^{-\frac{2i\nu}{\sqrt{f'_+g'_+}}}\biggr|^2.
\end{split}
\end{equation}
The final form of the probability in eq.(\ref{1.42}) takes the following form 
\begin{equation}\label{1.42a}
\begin{split}
P_\kappa&\cong\frac{4\pi\mathcal{G^2}\nu}{\sqrt{f'_+g'_+}\mathcal{B}^2\varpi^2}\left[1-\frac{4\gamma\nu\mathcal{e}}{f'_+\varpi\mathcal{B}}+\frac{g'_+\mathcal{e}^2\gamma^2}{f'_+\varpi^2\mathcal{B}^2}+\frac{4\nu\mathcal{e}\mu}{\mathcal{B}f'_+\varpi}\right]\\&\times\frac{1}{e^{\frac{4\pi\nu}{\sqrt{f'_+ g'_+}}}-1}~.
\end{split}
\end{equation}
where we have neglected $\mathcal{O}(\frac{\nu^2}{\varpi^2})$ order quantities in the parenthesis of the amplitude factor (as $\nu\ll \varpi$). 
We observe that in eq.(\ref{1.42}) the probability has Planck like factor which is independent of $\kappa$. In eq.(\ref{1.42a}), the term in the parenthesis can be expressed in the following form
\begin{equation}\label{1.42b}
\begin{split}
&1-\frac{4\gamma\nu\mathcal{e}}{f'_+\varpi\mathcal{B}}+\frac{g'_+\mathcal{e}^2\gamma^2}{f'_+\varpi^2\mathcal{B}^2}+\frac{4\nu\mathcal{e}\mu}{\mathcal{B}f'_+\varpi}\cong1-\frac{3\nu}{\mathcal{e}\varpi}\\&+\frac{\left[2{f'_+}^2g'_++\mathcal{e}^2f''_+g'_+-\mathcal{e}^2f'_+(g''_+-4\mu g'_+)\right]^2}{16\mathcal{e}^2\varpi^2g'_+{f'_+}^3}+\frac{\nu}{\varpi}\mathcal{k}
\end{split}
\end{equation}
%
%
%
where
\begin{equation}\label{1.43b}
\begin{split}
\mathcal{k}&=\frac{\mathcal{e}(-f''_+g'_++3f'_+g''_+)}{2{f'_+}^2g'_+}+(-2{f'_+}^2g'_++\mathcal{e}^2 f''_+g'_+
\\&+\mathcal{e}^2f'_+g''_+)\frac{\left[2{f'_+}^2g'_++\mathcal{e}^2 f''_+g'_+-\mathcal{e}^2f'_+(g''_+-4\mu g'_+)\right]^2}{16\mathcal{e}^3\varpi^2{g'_+}^2{f'_+}^5}~.
\end{split}
\end{equation}
If one neglects the $\mu$ term and the beyond NH approximation, the probability in eq.(\ref{1.42a}) takes the form
\begin{equation}\label{1.43}
P_{\kappa=0,NH}\cong\frac{4\pi\mathcal{G}^2\nu}{\sqrt{f'_+g'_+}\varpi^2}\left(1-\frac{3\nu}{\mathcal{e}\varpi}+\frac{f'_+g'_+}{4\mathcal{e}^2\varpi^2}\right)\dfrac{1}{e^{\frac{4\pi\nu}{\sqrt{f'_+g'_+}}}-1}.
\end{equation}
From eq.(\ref{1.42a}), we observe that the conformal symmetry plays no vital role in determining the form of the Planck factor indicating that the overall spectrum does not get affected even with the breaking of conformal symmetry. However, the amplitude gets modified in contrast to the case where there is conformal symmetry. In eq.(s)(\ref{1.42a},\ref{1.43}),  the frequency $\nu$ is observed by a distant observer. Therefore, this frequency $\nu=\nu_\infty$ can be expressed in terms of locally observed frequency ($\nu_\mathcal{o}$) as, $\nu_\infty=\nu_\mathcal{o}\sqrt{f(r)}\sim \nu_\mathcal{o}\sqrt{(r-r_+)f'(r_+)}=\frac{\nu_\mathcal{o}}{2a}\sqrt{f'_+g'_+}$. Substituting the form of $\nu_\infty$ in the Planck factors of eq.(s)(\ref{1.42},\ref{1.43}), we obtain the following form
\begin{equation}\label{1.44}
\begin{split}
\dfrac{1}{e^{\frac{4\pi\nu}{\sqrt{f'_+g'_+}}}-1}&\cong\frac{1}{e^{\frac{2\pi\nu_\mathcal{o}}{a}}-1}
\end{split}
\end{equation}
which is similar to the Planck factor for the case of a fixed atom and an accelerating mirror in a flat spacetime background. This result implies that the effect of the accelerating mirror on the field modes of the atom are similar to that of the gravitational effect of the black hole. Hence, we observe that for a generic black hole geometry the equivalence principle holds, and is independent of whether there is conformal symmetry or not.  The form of the excitation probability in terms of the acceleration $a$ and locally observed frequency $\nu_\mathcal{o}$ is given as follows
\begin{equation}\label{1.45}
\begin{split}
P_\kappa&=\frac{2\pi\mathcal{G}^2\nu_\mathcal{o}c}{a\varpi^2}\biggr[1-\frac{3\nu_\mathcal{o}c^2\sqrt{f'_+g'_+}}{2a\mathcal{e}\varpi}-\frac{\mathcal{e}\nu_\mathcal{o}c^2\sqrt{f'_+g'_+} f''_+}{2a{f'_+}^2\varpi}\\&+\frac{c^2\left[2{f'_+}^2g'_++\mathcal{e}^2f''_+g'_+-\mathcal{e}^2f'_+(g''_+-4\mu g'_+)\right]^2}{16\mathcal{e}^2\varpi^2g'_+{f'_+}^3}
\\&+\frac{3\mathcal{e}\nu_\mathcal{o}c^2 g''_+}{2a\sqrt{f'_+g'_+}\varpi}+\frac{\nu c^4\sqrt{f'_+g'_+}}{2a\varpi}(-2{f'_+}^2g'_++\mathcal{e}^2f''_+g'_+\\&+\mathcal{e}^2f'_+g''_+)\frac{\left[2{f'_+}^2g'_++\mathcal{e}^2f''_+g'_+-\mathcal{e}^2f'_+(g''_+-4\mu g'_+)\right]^2}{16\mathcal{e}^3\varpi^2{g'_+}^2{f'_+}^5}\biggr]\\&\times \frac{1}{e^{\frac{2\pi\nu_\mathcal{o}c}{a}}-1}
\end{split}
\end{equation}
where 
\begin{equation}\label{1.46}
\mu=\frac{\kappa}{1+\frac{\nu_\mathcal{o}^2c^2}{a^2}}~.
\end{equation}
In 4 dimensions, $\mu$ takes the following form (for $l=0$ and $\beta$=0 (as rest mass of the photon is zero))
\begin{equation}\label{1.46a}
\mu=\frac{\kappa}{1+\frac{4\nu^2}{f'_+g'_+c^2}}\cong \frac{f''_+}{8f'_+}+\frac{g''_+}{8g'_+}+\frac{1}{r_+}.
\end{equation}
Now from eq.(\ref{1.32a}), we can write that (for $\mathcal{l}=0$)
\begin{equation}\label{1.46b}
\left(\frac{dr}{d\tau}\right)^2=\frac{g(r)}{f(r)}\mathcal{e}^2-g(r)~.
\end{equation}
In the $r\rightarrow\infty$ limit, $f(r\rightarrow\infty)=g(r\rightarrow\infty)=1$~\footnote{Standard black hole solutions satisfy this condition.}. Thus, eq.(\ref{1.46b}) reduces to the following form
\begin{equation}\label{1.46c}
\left(\frac{dr}{d\tau}\right)^2=\mathcal{e}^2-1~.
\end{equation}
For $\mathcal{e}=1$, we observe from $eq.(\ref{1.46c})$ that $\frac{dr}{d\tau}=0$ (in the $r=\infty$ limit). Hence, $\mathcal{e}=1$ corresponds to the rest mass energy of the atom at $r=\infty$. But since $\mathcal{e}$ is a conserved quantity, hence it corresponds to the rest mass energy of the atom. From now on we shall consider the case $\mathcal{e}=1$.
Using eq.(\ref{1.46a}) and $\mathcal{e}=1$, the probability in eq.(\ref{1.45}) takes the following form
\begin{equation}\label{1.46d}
\begin{split}
P_\kappa&=\frac{2\pi\mathcal{G}^2\nu_\mathcal{o}c}{a\varpi^2}\biggr[1-\frac{3\nu_\mathcal{o}c^2\sqrt{f'_+g'_+}}{2a\varpi}-\frac{\nu_\mathcal{o}c^2\sqrt{f'_+g'_+} f''_+}{2a{f'_+}^2\varpi}\\&+\frac{c^2\left[8f'_+g'_++4{f'_+}^2g'_+r_++3f''_+g'_+r_+-f'_+g''_+r_+\right]^2}{64\varpi^2r_+^2g'_+{f'_+}^3}
\\&+\frac{3\nu_\mathcal{o}c^2 g''_+}{2a\sqrt{f'_+g'_+}\varpi}+\frac{\nu c^4\sqrt{f'_+g'_+}}{2a\varpi}(-2{f'_+}^2g'_++f''_+g'_+\\&+f'_+g''_+)\frac{\left[8f'_+g'_++4{f'_+}^2g'_+r_++3f''_+g'_+r_+-f'_+g''_+r_+\right]^2}{64\varpi^2r_+^2{g'_+}^2{f'_+}^5}\biggr]\\&\times \frac{1}{e^{\frac{2\pi\nu_\mathcal{o}c}{a}}-1}~.
\end{split}
\end{equation}
Interestingly, even though the Planck factor remains unaffected, the coefficient in front of the excitation probability gets changed and has dependence on the acceleration term. This is the consequence of taking terms beyond the near horizon approximation in the analysis which breaks the conformal symmetry. Note that the case $\mathcal{e}>1$ corresponds to the fact that the atom is not at rest at $r=\infty$ since from eq.(\ref{1.46c}), we see that $\frac{dr}{d\tau}>0$ at $r=\infty$.
\section{GHS black hole and the HBAR entropy}
In this section, we will try to calculate the \textit{horizon brightened acceleration radiation} entropy (HBAR)  entropy for the case of a GHS black hole with its metric given in eq.(s)(\ref{1.2})-(\ref{1.4}). The horizon radius for a GHS black hole is at $r_+=\frac{2GM}{c^2}e^{\phi_0}$. We now consider a set of two level atoms falling into the event horizon of the GHS black hole at a rate $\mathcal{k}$. Now if $\Delta N$ is the number of atoms falling in a time interval $\Delta t$, then we can write
\begin{equation}\label{1.47}
\frac{\Delta N}{\Delta t}=\mathcal{k}~.
\end{equation} 
If $\delta \rho_j$ is the change in the density matrix due to the $j$-th atom, then for $\Delta N$ number of atoms, the total change in the density matrix is given by (assuming $\delta\rho_i=\delta\rho_j=\delta\rho~;~\forall i,j$ )
\begin{equation}\label{1.48}
\begin{split}
\Delta\rho&=\sum\limits_j \delta\rho_j=\Delta N \delta\rho\\
\implies\frac{\Delta\rho}{\Delta t}&=\mathcal{k}\delta\rho~.
\end{split}
\end{equation}
Now if $\Gamma_{exc}=\mathcal{k}\mathcal{P}_{exc}$ and $\Gamma_{abs}=\mathcal{k}\mathcal{P}_{abs}$ give the excitation and absorption rates for a GHS black hole (with $\mathcal{P}_{exc}$ and $\mathcal{P}_{abs}$ describing the excitation and absorption probabilities) then the Lindblad master equation reads\cite{Lindblad}
\begin{equation}\label{1.49}
\begin{split}
\dot{\rho}&=-\frac{\Gamma_{abs}}{2}\left[\rho b^\dagger b+b^\dagger b\rho-2b\rho b^\dagger\right]\\
&~~~-\frac{\Gamma_{exc}}{2}\left[\rho b b^\dagger+b b^\dagger\rho-2 b^\dagger\rho b\right]~.
\end{split}
\end{equation}
With respect to some arbitrary state $\left|n\right>$, the expectation value of eq.(\ref{1.49}) is given as follows
\begin{equation}\label{1.50}
\begin{split}
\dot{\rho}_{n,n}=&-\Gamma_{abs}\left(n\rho_{n,n}-(n+1)\rho_{n+1,n+1}\right)\\
&-\Gamma_{exc}\left((n+1)\rho_{n,n}-n\rho_{n-1,n-1}\right)~.
\end{split}
\end{equation} 
The steady state solution of eq.(\ref{1.50}) is given as follows\cite{OTM}
\begin{equation}\label{1.51}
\rho^{\mathcal{S}}_{n,n}=\left(\frac{\Gamma_{exc}}{\Gamma_{abs}}\right)^n\left(1-\frac{\Gamma_{exc}}{\Gamma_{abs}}\right) 
\end{equation}
where for a GHS black hole
\begin{equation}\label{1.52}
\begin{split}
\frac{\Gamma_{exc}}{\Gamma_{abs}}&\cong\left(1-\frac{6\nu}{\varpi}+\frac{2\nu}{\varpi}\mathcal{k}\right)e^{-\frac{4\pi\nu}{c\sqrt{f'_+g'_+}}}\\
&\cong e^{-\frac{8\pi\nu G M e^{\phi_0}}{c^3}} \biggr[1-\frac{10\nu}{\varpi}+\frac{3e^{2\phi_0}Q^2\nu}{2GM^2\varpi\pi\varepsilon_0}-\frac{3c^6e^{-2\phi_0}\nu}{G^2M^2\varpi^3}\\
&+\frac{11c^6Q^2\nu}{8G^3M^4\varpi^3\pi\varepsilon_0}-\frac{51c^6e^{2\phi_0}Q^4\nu}{256G^4M^6\varpi^3\pi^2\varepsilon_0^2}\\&+\frac{9c^6e^{4\phi_0}Q^6\nu}{1024G^5M^8\varpi^3\pi^3\varepsilon_0^3}\biggr]
\end{split}
\end{equation}
with  $\varepsilon_0$ being the permittivity of free space. To obtain eq.(\ref{1.52}), we have considered the following assumptions; $\nu\ll\varpi$, $f'_+c,~g'_+c,\sqrt{f''_+}c,\sqrt{g''_+}c\ll\varpi$, and $e^{\phi_0}\ll 1$~.
The von Neumann entropy and its rate of change for the system are given by\cite{Fulling2}
\begin{align}
S_\rho&=-k_B\sum\limits_{n,\nu}\rho_{n,n}\ln\rho_{n,n}~,\label{1.53}\\
\dot{S}_\rho&=-k_B\sum\limits_{n,\nu}\dot{\rho}_{n,n}\ln\rho_{n,n}\cong-k_B\sum\limits_{n,\nu}\dot{\rho}_{n,n}\ln\rho^{\mathcal{S}}_{n,n}\label{1.54}
\end{align}
where in case of the rate of change, we have replaced $\rho_{n,n}$ by the steady state solution $\rho_{n,n}^{\mathcal{S}}$. Using the form of $\rho_{n,n}^{\mathcal{S}}$ from eq.(s)(\ref{1.51},\ref{1.52}), we obtain the analytical form of $S_\rho$ as follows
\begin{equation}\label{1.55}
\begin{split}
\dot{S}_\rho\cong&-k_B\sum\limits_{n,\nu}n\dot{\rho}_{n,n}\biggr(-\frac{8\pi\nu G Me^{\phi_0}}{c^3}\\&+\ln\biggr[1-\frac{10\nu}{\varpi}+\frac{3e^{2\phi_0}Q^2\nu}{2GM^2\varpi\pi\varepsilon_0}-\frac{3c^6e^{-2\phi_0}\nu}{G^2M^2\varpi^3}\\
&+\frac{11c^6Q^2\nu}{8G^3M^4\varpi^3\pi\varepsilon_0}-\frac{51c^6e^{2\phi_0}Q^4\nu}{256G^4M^6\varpi^3\pi^2\varepsilon_0^2}\\&+\frac{9c^6e^{4\phi_0}Q^6\nu}{1024G^5M^8\varpi^3\pi^3\varepsilon_0^3}\biggr]\biggr)\\
\cong&\biggr(\frac{8\pi GMk_Be^{\phi_0}}{c^3}+\frac{10k_B}{\varpi}-\frac{3k_Be^{2\phi_0}Q^2}{2GM^2\varpi\pi\varepsilon_0}\\
&+\frac{3k_Bc^6e^{-2\phi_0}}{G^2M^2\varpi^3}-\frac{11k_Bc^6Q^2}{8G^3M^4\varpi^3\pi\varepsilon_0}+\frac{51k_Bc^6e^{2\phi_0}Q^4}{256G^4M^6\varpi^3\pi^2\varepsilon_0^2}\\&-\frac{9k_Bc^6e^{4\phi_0}Q^6}{1024G^5M^8\varpi^3\pi^3\varepsilon_0^3}\biggr]\biggr)\sum\limits_{\nu}\dot{\bar{n}}_\nu\nu
\end{split}
\end{equation}
where $\bar{n}_\nu$ is the flux due to photons generated from two-level atoms falling in the event horizon of the black hole. The total loss of energy due to emitted photons is given as $\hbar\sum\limits_\nu \dot{\bar{n}}_\nu\nu=\dot{m}_pc^2$. If $M$ is the mass of the black hole, then the rate of change of mass can be expressed as $\dot{M}=\dot{m}_p+\dot{m}_{atom}$. Here, $\dot{m}_p$ is the rate of change of the rest mass of the black hole due to emitting photons\cite{Fulling2}. The rate of change of area of the black hole due to the emitted photons can be expressed as follows
\begin{equation}\label{1.56}
\dot{A}_p=\frac{32\pi G^2M\dot{m}_p}{c^4}e^{2\phi_0}~.
\end{equation}  
This result can be interpreted in the following way. Before entering the event horizon (before contributing to the mass of the black hole) the atom emits radiation. Hence, the black hole entropy associated with HBAR radiation from an atom and that associated with an atom can be separated in time. Hence, we can say that when no atoms are falling in the black hole, $A_p$ can be considered to be equal to the area of the entire black hole. The final form of $\dot{S}_\rho$ in terms of $A_p$ can be written as follows 
\begin{equation}\label{1.57}
\begin{split}
\dot{S}_\rho&=\frac{d}{dt}\biggr[\frac{k_Bc^3e^{-\phi_0}}{4\hbar G}A_p+\frac{5k_Bc^4e^{-\phi_0}}{2\hbar\varpi\sqrt{\pi}G}A^{\frac{1}{2}}_p+\frac{6k_BQ^2e^{3\phi_0}}{\hbar\varpi\sqrt{\pi}\varepsilon_0}A^{-\frac{1}{2}}_p\\&-\frac{12c^6k_B\sqrt{\pi}e^{-\phi_0}}{\hbar G\varpi^3}A^{-\frac{1}{2}}_p+\frac{88c^2k_B\sqrt{\pi}Q^2e^{3\phi_0}}{3\hbar\varpi^3\varepsilon_0}A_p^{-\frac{3}{2}}
\\&-\frac{204k_B\sqrt{\pi}GQ^2e^{7\phi_0}}{5\hbar\varpi^3\varepsilon_0^2c^2}A^{-\frac{5}{2}}_p+\frac{144k_b\sqrt{\pi}G^2Q^6e^{11\phi_0}}{\hbar\varpi^3\varepsilon_0^3c^6}A_p^{-\frac{7}{2}}\biggr].
\end{split}
\end{equation}
If we now consider the near horizon approximation only then the probability in eq.(\ref{1.57}) takes the form
\begin{equation}\label{1.58}
\dot{S}_\rho=\frac{d}{dt}\biggr(\frac{k_Bc^3e^{-\phi_0}}{4\hbar G}A_p+\frac{3k_Bc^4e^{-\phi_0}}{2\hbar\varpi\sqrt{\pi}G}A^{\frac{1}{2}}_p\biggr)~.
\end{equation}
This is also a new result in our paper. Comparing eq.(\ref{1.57}) with eq.(\ref{1.58}), we observe that the rate of change of the entropy misses out important corrections in terms of the area $A_p$ while considering conformal symmetry. Interestingly, we observe that in eq.(\ref{1.57}) the third and fourth terms are similar to that of quantum gravity corrections to the HBAR entropy\cite{OTM} observed for the case of a quantum corrected black hole. We also observe that the coefficient of the second term in eq.(\ref{1.58}) gets modified in eq.(\ref{1.57}) and three additional higher order corrections (in terms of area) appear due to the consideration of the beyond near horizon approximation.
\section{Conclusion}
In this work, we have considered a two-level atom falling into the event horizon of a large class of static spherically symmetric black holes covered by a mirror to shield infalling atoms from interacting with the Hawking radiation. This set up is identical to that of a Boulware vacuum\cite{Boulware}. For an external observer, the initial state of the field appears vacuum-like due to the mirror. We investigated the scalar field solution in the beyond near horizon approximation which leads to neglecting conformal symmetry in the equations. The analysis is important as it captures the effect of going beyond the near horizon approximation. The HBAR entropy for a GHS black hole in this context is also investigated. It is observed that the Planck factor obtained in the excitation probability does not get affected due to the beyond near horizon approximation. However, we see significant modifications in the coefficient of the transition probability. Interestingly, we also find that the equivalence relation holds in the case of such a large class of spherically symmetric and static black hole geometries even when terms breaking the conformal invariance are included in the analysis. It indicates the fact that conformal invariance does not play a vital role in the overall understanding of the equivalence relation. For the next part of our analysis, we have tried to motivate this large class of static and spherically symmetric black hole geometries by considering the example of the Garfinkle-Horowitz-Strominger black hole which arises in low energy string theory. We have calculated the HBAR entropy for both the conformal case and the non-conformal case. We observe that in the case of the beyond near horizon approximation, the HBAR entropy consists of a term similar to Bekenstein-Hawking entropy (multiplied with an exponential term) along with square root, inverse square root corrections and higher orders of inverse corrections in terms of the area $A_p$. Surprisingly, the inverse square root correction does not occur when we consider the conformal symmetry case along with near horizon approximation. We also observe a different coefficient for the square root correction term in the HBAR entropy. It is to be noted that this inverse square root correction is similar to a quantum gravity correction in the HBAR entropy of a quantum corrected black hole\cite{OTM}. 
\section*{Acknowledgement}
We thank the referee for useful and constructive comments.

\end{document}